\documentclass{iaus} 
\usepackage{amssymb,amsmath}
\usepackage{graphicx}

\newcommand{\oversim}[2]{\protect{\mbox{\lower0.5ex\vbox{%
   \baselineskip=0pt\lineskip=0.2ex
   \ialign{$\mathsurround=0pt #1\hfil##\hfil$\crcr#2\crcr\sim\crcr}}}}} 
\newcommand{\simgreat}{\mbox{$\,\mathrel{\mathpalette\oversim>}\,$}} 
\newcommand{\simless} {\mbox{$\,\mathrel{\mathpalette\oversim<}\,$}} 
\title[Binary and stellar populations] 
{The universality hypothesis: binary and stellar populations in star clusters and galaxies}
\author[Pavel Kroupa]   
{Pavel Kroupa}

\affiliation{Argelander-Institut f\"ur Astronomie, Universit\"at Bonn,
  Auf dem H\"ugel 71, D-53121 Bonn, Germany\\ email: {\tt
    pavel@astro.uni-bonn.de}}

\pubyear{2010}
\volume{270}  
\pagerange{xxx--xxx}
\jname{Computational Star Formation}
\editors{J. Alves, B. Elmegreen, J. Girart, V. Trimble,eds.}
\begin{document}

\maketitle

\begin{abstract}
It is possible to extract, from the observations, distribution
functions of the birth dynamical properties of a stellar population,
and to also infer that these are quite invariant to the physical
conditions of star formation. The most famous example is the stellar
IMF, and the initial binary population (IBP) seems to follow suit. A
compact mathematical formulation of the IBP can be derived from the
data. It has three broad parts: the IBP of the dominant stellar
population ($0.08-2\,M_\odot$), the IBP of the more-massive stars and
the IBP of brown dwarfs. These three mass regimes correspond to
different physical regimes of star formation but not to structure in
the IMF. With this formulation of the IBP it becomes possible to
synthesise the stellar-population of whole galaxies.  
\end{abstract}

\firstsection 

\section{Introduction}
\label{sec:introd}

\noindent
The fundamental dynamical properties of stellar populations are the
masses of the stars and their correlation in multiple stellar
systems. The distribution of stellar masses at birth, the IMF, is
rather well constrained and has (surprisingly) been found to be
invariant despite theoretical models predicting systematic variation
for example of the mean stellar mass or even of the minimum mass with the
physical conditions of star formation. This problematical issue has
been discussed at some length by \cite{K08}, where the {\sc IMF
  Universality Hypothesis} is stated.

Equivalently, the question may be raised whether the other
distribution functions characterising a stellar population, namely the
distribution functions of binary systems, are just as invariant. If
this were the case then it would have important bearings on the theory
of star formation as the fragmentation length-scale may then not
depend much on the physical conditions of the molecular cloud core. A
change in the properties of the binary-star distribution functions
with mass scale, if found, would yield important clues to the
fragmentation and angular momentum re-distribution processes during
star formation.

The three important distribution functions describing the initial
binary population (IBP) are the distribution of periods ($P$, here
always in days), or equivalently of semi-major axes ($a$, in AU), the
distribution of mass-ratios ($q=m_2/m_1 \le 1$) and the distribution
of orbital eccentricities ($e$). These are related by Kepler's third
law: $a^3/P_{\rm yr}^2 = m_1+m_2$, where $P_{\rm yr}$ is the orbital
period in years ($P=365.25\,P_{\rm yr}$) and $m_1, m_2$ are the
primary- and secondary-star masses in $M_\odot$. Because the periods
of binary stars range over many orders of magnitude the shorthand
$lP\equiv {\rm log}_{10}P$ is used throughout this text.


\section{Star formation and the initial binary population (IBP)}
\label{sec:sf}

\noindent
Observations have shown that the star-formation process is
intrinsically linked to the production of binary stars. Indeed, binary
stars must be the dominant formation channel because the observed
multiplicity fraction (the number of multiple systems divided by the
number of sources in the survey) is indistinguishably high among old
metal-poor (\cite{Carneyetal05}) and among thin-disk main-sequence
stars \newline (\cite{DM91}), and is near unity for pre-main sequence stars and
proto-stars (\cite{D99, Connelley08}). If, on the other hand,
higher-order multiple systems were a major outcome of late-type star
formation, then the dynamical decay of these on a time-scale
$<10^5\,$yr would pollute the pre-main-sequence stellar population
with single stars which are not observed in large numbers.  Indeed,
this is evidently the case for massive stars ($m\simgreat\,{\rm
  few}\,M_\odot$) which appear to form preferentially in binary-rich
dense cores of populous embedded clusters which rapidly decay
dynamically by ejecting massive stars (\cite{CP92, PAK10}).  Thus,
according to the {\sc Binary-Star Conjecture or Theorem} (\cite{K08})
the vast number of stars form as binaries, while non-hierarchical
higher order multiple systems cannot be a significant outcome of
late-type star formation.

The formation of binary systems remains an essentially unsolved
problem theoretically. \cite{Fisher04} shows analytically that
isolated turbulent cloud cores can produce an unquantifiable fraction
of binary systems with the very wide range of orbital periods as
observed. But direct cloud collapse calculations are very limited in
predicting binary-star properties owing to the severe computational
difficulties of treating the magneto-hydrodynamics together with
correct radiation transfer and evolving atomic and molecular opacities
during collapse.

The currently most advanced hydrodynamical simulations have been
reported by \newline \cite{MB10}. They allow a turbulent SPH cloud to collapse
forming a substantial cluster of 1253~stars and brown dwarfs amounting
to $191\,M_\odot$. The cluster has a half-mass radius of about 0.05~pc
and contains a very substantial binary and higher-order multiple
stellar population with a large spread of semi-major axes but peaking
at a few~AU. After dynamical evolution with or without expulsion of
the residual gas the distribution of orbits ends up being quite
strongly peaked at a few~AU with a significant deficit of orbits with
$a>10\,$AU, and with a deficit of systems with a mass ratio $q<0.8$,
when compared to the main-sequence population (their fig.~11). This
state-of-the art computation therewith confirms the above stated issue
that it remains a significant challenge for star-formation theory to
account for the Gaussian-type distribution of $a$ spanning
$10^{-1}-10^5\,$AU as for Galactic-field binaries.  One essential
aspect which is still missing from such computations is stellar
feedback which starts heating the cloud as soon as the first proto
stars appear. These heating sources are likely to counter the
gravitational collapse such that in reality the extreme densities are
not achieved allowing a much larger fraction of wide binaries to
survive.

More general theoretical considerations suggest that star-formation in
dense clusters ought to have a tendency towards a {\it lower} binary
proportion in warmer molecular clouds (i.e. in cluster-forming cores)
because of the reduction of available phase-space for binary-star
formation with increasing temperature (\cite{DS94}, hereinafter DS).
On the other hand, an {\it enhanced} binary proportion for orbital
periods $lP \simless 5.6$ may be expected in dense clusters due to the
stimulation of binary formation through tidal shear (\cite{HBB01}),
thus possibly compensating the DS effect.  The initial period
distribution function (IPF) may thus appear similar in dense and
sparse clusters, apart from deviations at long periods due to
encounters and the cluster tidal field.

The multiple-star population in the Galactic field is build-up by
star-forming events in star-clusters or groups containing from a dozen
to possibly millions of stars. Indeed, a certain but presently not
well known fraction of stars form in small--$N$ systems that typically
have a size $\approx 100\,$AU, and the dynamical decay of these is
likely to affect the final distribution of $P\simless10^5$ binaries
(\cite{SD98}), giving rise to non-uniform jet activity
(\cite{R00}). But again, quantification of their properties is
next-to-impossible given the neglect of the hydrodynamical
component. But, by the {\sc Binary-Star Conjecture} above, the binary
formation channel must be vastly dominating over the formation of
non-hierarchical higher-order multiples.

\section{The IBP universality hypothesis}
\label{sec:sfuniv}

\noindent
It is hoped that star-forming simulations will allow essential
insights into reproducing the stellar population stemming from an
individual modest star-forming event. But given the computational
complexity, synthesizing for example the binary population in a
massive star cluster or of a whole galaxy or even parts thereof such
as the solar neighbourhood are not possible.  Therewith it becomes
rather apparent that current theory has no predictive power concerning
the binary properties of stellar populations in different
environments.

However, by proposing that initial distribution functions of the
binary-star properties (see end of \S~\ref{sec:introd}) exist, that
is, that there exists an outcome of the isolated (i.e. low-density,
such as in Taurus-Auriga) star-formation process that can be
quantified in terms of an IBP, we would be put into the situation of
being able to synthesise populations. This would become feasible if it
is understood how this IBP is affected by physical processes that are
inherent to a binary system and that are due to stellar-dynamical
encounters in denser star-forming regions.

In fact, the invariance of the IMF must be implying an insensitivity
of the star-formation outcome to physical conditions. The IMF being
invariant constitutes a statistical statement on one of the birth
dynamical properties of stars (namely their distribution of
masses). So, since both the IMF and the IBP are the result of the same
(star-formation) process, and since the IMF is a result of this
process ``one level deeper down'' than the IBP, it is quite natural to
suggest that the formal mathematical distribution function of all of
the birth dynamical properties of stars are invariant.  Thus the
following hypothesis follows:

\vspace{1mm}

\centerline{ \fbox{\parbox{8.5cm}{
{\sc 
The Star-Formation Universality
        Hypothesis:}
\newline 
IMF universality $\Longleftrightarrow$
      IBP universality.
\label{hyp_pk:univ}}}}

\section{Inferring the IBP for $m<2\,M_\odot$ stars}
\label{sec:IBP}

\begin{figure}[b]
\vspace*{-0.2cm}
\begin{center}
 \includegraphics[width=4in]{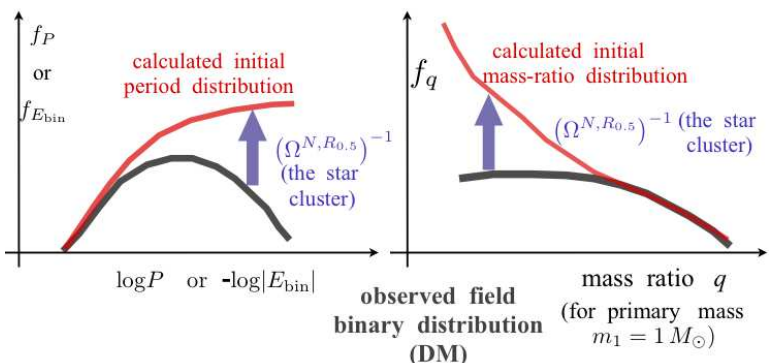} 
 \vspace*{-0.3 cm}
 \caption{Schematic of how the observed Galactic-field main-sequence
   (thick black curves, \cite{DM91}) period or binding energy
   distribution ({\bf left panel}) and the main-sequence mass-ratio
   distribution ({\bf right panel}) are back-computed (upwards
   pointing arrows) with the stellar-dynamical operator
   $\Omega^{M_{\rm ecl},R_{0.5}}$ to yield an estimate of the initial
   (thin red curves) period and mass-ratio distributions. Note that
   the physical reality of $\Omega^{M_{\rm ecl}, R_{0.5}}$ is
   established if both $f_P$ and $f_q$ become consistent with the
   pre-main sequence data for one and the same $\Omega^{M_{\rm ecl},
     R_{0.5}}$. The eccentricity distribution is not affected by
   $\Omega^{M_{\rm ecl},R_{0.5}}$.}
   \label{fig:krcid}
\end{center}
\end{figure}

\noindent
The challenge of inferring the period-, mass-ratio- and
eccentricity-distribution functions characterising the IBP can be
formulated as follows: assuming the {\sc Star-Formation Universality
  Hypothesis} to be valid and using observational constraints on the
pre-main sequence binary population, the observed main-sequence
Galactic-field binary-star distribution function must be corrected for
the dynamical processes acting in the birth-groups or birth-clusters
of stars as these emerge into the Galactic field. Fig.~\ref{fig:krcid}
visualises this idea.

These dynamical processes are well understood and are detailed in
\cite{K08}: Energy arguments imply the {\sc Heggie-Hills Law}
according to which the wide-binaries are disrupted preferentially
compared to short-period binaries and tight binaries typically become
tighter.  The boundary between the short and long-period binaries
depends on the velocity dispersion in the birth population. At the
same time, binaries with a small mass ratio (small $q$) are also
preferentially disrupted. The tightening of tight binaries implies
that such systems and individual stars interacting with them can be
ejected from the cluster.  Pre-main-sequence {\sc Eigenevolution}
evolves the short-period binaries within a time-scale of
few~$10^5\,$yr due to system-internal dissipative processes such as
tidal-circularisation, primary-star-disk--secondary-star interactions
and disk--disk interactions.

The dynamical disruption of binaries induced through the birth cluster
and characterised by the stellar-dynamical operator $\Omega^{M_{\rm
    ecl},R_{0.5}}$, and the mass-ratio, eccentricity and period
evolution induced through pre-main sequence eigenevolution, can both
be calculated (see \cite{K08} for details).  $M_{\rm ecl}$ is the
stellar mass of the embedded cluster with half-mass radius $R_{0.5}$.
Note that this ``embedded cluster'' does not need to be a cluster
which survives the first few~$10\,$Myr as a bound star cluster. It can
readily be taken to include sparse star-forming aggregates of stars as
observed e.g. in the about 0.5~pc radius sub-groups of a dozen
pre-main sequence stars in Taurus-Auriga.

The above two-stage transformation of the formal mathematical birth
distribution functions ($f_P, f_q, f_e$) to the final distribution
function (e.g. of the dispersed cluster) can be written as
\begin{equation}
{\cal D}_{\rm outcome}(lP, e, q : m_1) = \Omega^{M_{\rm ecl},R_{0.5}} \left[ \Omega_{\rm
    eigenevol}\left[ {\cal D}_{\rm birth}(lP, e, q : m_1)\right] \right].
\label{eq:kromega}
\end{equation}
${\cal D}_{\rm birth}=f_{P}(lP)\,f_q(q)\, f_e(e)$ is the birth
distribution function taking the birth period, mass-ratio and
eccentricity distribution functions to be separable, that is, the
birth parameters $P_b, q_b, e_b$ are not correlated.

A simple theoretical treatment of eigenevolution which is based on
pre-main sequence tidal-circularisation theory has been shown to quite
nicely account for the correlations between eccentricity, mass-ratio
and period for short-period ($P\simless 10^3\,$d) binaries. The
operator $\Omega_{\rm eigenevol}$ generates the correlations between
period, eccentricity and mass ratio.

The stellar-dynamical operator, $\Omega^{M_{\rm ecl},R_{0.5}}$, is
given by an Nbody star-cluster, and can be envisioned as a
transformation of the number of binaries in different binding-energy
bins such that the reduction is largest in the most weakly-bound
binaries. The tightening of binaries can increase the fraction of
short-period binaries in a population.

Putting this together, the birth binary population is defined by the
following formal mathematical rules:

\vspace{2mm}

\centerline{ \fbox{\parbox{13.5cm}{\vspace{1mm}
{\sc The birth binary population (BBP):\vspace{1mm}}
\begin{itemize}
\item random pairing from the canonical IMF for $0.08\simless
  m/M_\odot \simless 2$;
\item thermal eccentricity distribution of eccentricities, $f_e(e)=2\,e$;
\item the period distribution function
\begin{equation}
f_{P,birth}=\eta \, {lP-lP_{\rm min} \over \delta + \left(lP -
  lP_{\rm min}\right)^2},
\label{eq:fPbirth}
\end{equation}
where $\eta=2.5, \delta=45, lP_{\rm min}=1$ and $\int_{lP_{\rm
    min}}^{lP_{\rm max}} f_{P,birth} \, dlP = 1$ such that the birth
binary fraction is unity ($lP_{\rm max}=8.43$).
\vspace{1mm}
\end{itemize}
\label{def_pk:bbp}}}}

\vspace{2mm}

\noindent Details are provided in \cite{K08}. Note that customarily
the ``initial binary population'' (IBP) derives from the BBP after
pre-main sequence eigenevolution. Given that pre-main sequence
eigenevolution occurs mostly for short-period binaries on a time-scale
of $10^5\,$yr within the system, it is expected to be approximately
universal. The {\sc Star Formation Universality Hypothesis} thus
remains valid.

\begin{figure}[b]
\begin{center}
 \includegraphics[width=4in]{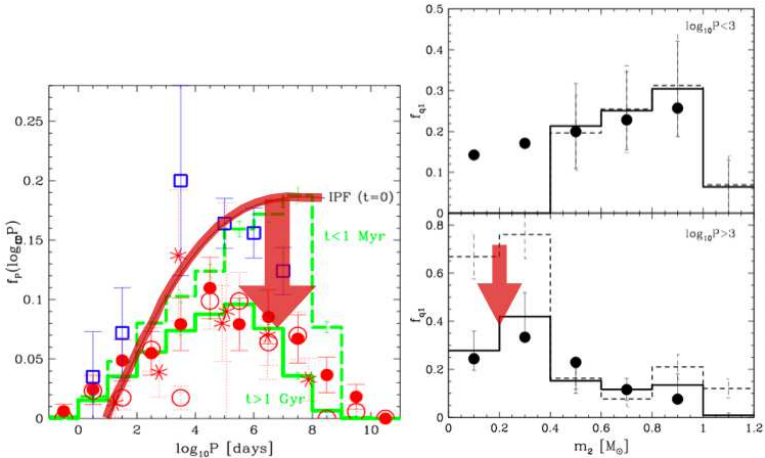} 
 \caption{ {\bf Left panel:} The transformation of the birth period
   distribution function (BPF, Eq.~\ref{eq:fPbirth}, thick red curve
   with $\eta=2.5, \delta=45, lP_{\rm min}=1$, \cite{K2}$=$K2) first
   to the eigenevolved ``initial'' period distribution function (IPF,
   dashed green histogram which can be described by
   Eq.~\ref{eq:fPbirth} with $\eta\approx3.5, \delta \approx 100,
   lP_{\rm min}=0$, \cite{K1}$=$K1). The green solid histogram is the
   final PF after $\Omega^{130\,M_\odot, 0.8\,{\rm pc}}$ acts on the
   IBP. The solid dots, open circles and stars are G-, K-, and M-dwarf
   binaries, and the open squares are pre-main sequence systems (see
   \cite{K08} for references). The normalisation of this plot is such
   that each period bin contains the fraction of binary orbits in the
   whole sample of stars plus binaries.  {\bf Right panel:} The
   transformation of the {\it initial} mass-ratio distribution for
   primary masses $m_1\approx1\,M_\odot$ (dashed histogram) to the
   final mass-ratio distribution after the population emerges from its
   star cluster (solid histogram, from K2). The upper panel is for
   short-period binaries which are not affected by
   $\Omega^{130\,M_\odot, 0.8\,{\rm pc}}$, while the bottom panel
   shows the long-period binaries. Note that the {\it birth}
   mass-ratio distribution, which results from random pairing from the
   IMF, is given by the dashed histogram in the lower panel, while
   eigenevolution transforms this distribution to the dashed histogram
   evident in the upper panel. That pre-main sequence,
   i.e. dynamically unevolved, mass-ratios are consistent with random
   paring from the IMF has been found by \cite{WLK01}.  }
   \label{fig:krPqevol}
\end{center}
\end{figure}

\begin{figure}[b]
\begin{center}
 \includegraphics[width=5in]{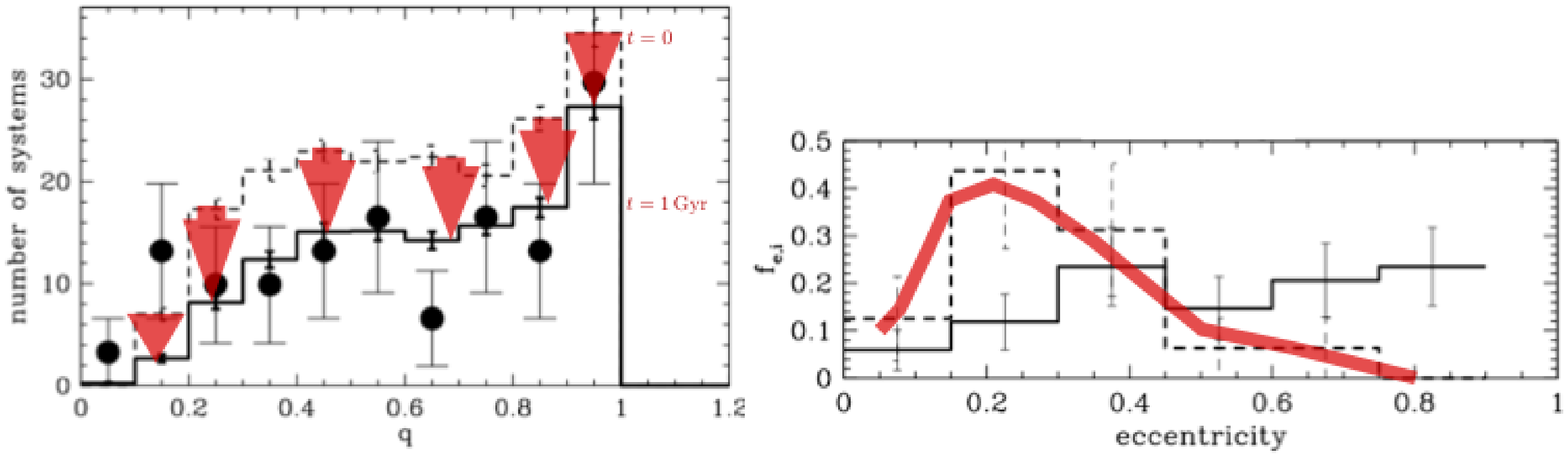} 
 \caption{{\bf Left panel:} The transformation of the overall {\it
     initial} mass-ratio distribution for $0.8 < m_1/M_\odot < 1$
   (dashed histogram) to the final mass-ratio distribution after the
   population emerges from its star cluster (solid histogram, from
   \cite{K08}). The $q=1$ peak results from pre-main sequence
   eigenevolution.  The solid circles are observational data by
   \cite{RG97}.  {\bf Right panel:} The eccentricity distribution for
   short-period ($lP<3$, dashed histogram) and long period ($lP>3$,
   solid histogram) binaries. The thick red solid curve visualises the
   bell-shaped distribution of short-period orbits which results from
   the thermal distribution (the solid histogram) after
   eigenevolution. Stellar-dynamical encounters, i.e. $\Omega^{M_{\rm
       ecl},R_{0.5}}$ have no effect on the thermal eccentricity
   distribution (it is an invariant to $\Omega^{M_{\rm ecl},
     R_{0.5}}$).}
   \label{fig:krqe}
\end{center}
\end{figure}

Passing the above easily generated {\it birth} distributions through
eigenevolution and then letting the stellar-dynamical operator
$\Omega^{M_{\rm ecl},R_{0.5}}$ act on this resulting {\it initial}
distribution leads to the Galactic field population as observed. This
is shown in Figs~\ref{fig:krPqevol}--\ref{fig:krqe} for the case
$M_{\rm ecl}=130\,M_\odot, R_{0.5}=0.8\,$pc, where the arrows indicate
the action of $\Omega^{M_{\rm ecl},R_{0.5}}$. These results are valid
for long-lived cluster models without gas. Short-lived embedded
clusters would have a smaller $R_{0.5}$ to give a {\it
  dynamically-equivalent} $\Omega^{M_{\rm ecl},R_{0.5}}$ (\cite{K1}).
Note that in Fig.~\ref{fig:krPqevol} the normalisation of the
distribution functions follows the custom
\begin{equation}
f = {N_{\rm orbits} \over  N_{\rm sys}},
\label{eq:krnorm}
\end{equation}
where $N_{\rm orbits}$ is the number of binary-star orbits found in a
sample of $N_{\rm sys}=N_{\rm orbits}+N_{\rm singles}$ systems or
  sources, and the $N_{\rm orbits}$ may be the total number of
  binaries (for the total binary fraction in a population), or just
  the number of orbits in an $lP$ interval yielding $f_P$.

\section{The IBP for massive stars ($m > 2\,M_\odot$) and for brown dwarfs}
\label{sec:IBP2}

\noindent
A seminal argument suggesting that massive stars form in regions of
high density void of low-mass stars but preferentially in tight
binaries with similar component masses has been provided by
\cite{CP92}. They reach this conclusion on the basis of the
distribution of OB runaway stars finding these conditions to be
necessary. In Bonn, Seungkyung Oh is building on these semi-analytical
results by performing direct Nbody calculations of realistic young
clusters and testing different pairing rules for massive stars in
initially mass-segregated and unsegregated clusters. The general
result is that massive stars indeed need high densities and tight
binaries with mass-ratios nearby unity in order to account for their
spatial distribution around star forming regions.

Brown dwarfs (BDs) form an altogether distinct population from stars
as is deduced by \cite{TK08} on the basis of the observed different
pairing properties of brown dwarfs and very-low-mass stars on the one
hand side, and stars on the other hand side. The most famous evidence
for this comes from the brown-dwarf desert: while stars pair up
irrespectively of their masses (e.g. G-dwarfs typically have M-dwarf
companions), BDs are exempt from participating (there are exceptions,
but such exceptions are a natural outcome of stellar-dynamical
processes). BDs also have a very different semi-major axis
distribution, by being limited to $a\simless 20\,$AU, while for stars
$a$ ranges over five orders of magnitude. Also, BDs have a small
binary fraction $f\approx 0.15$ with a bias towards similar companion
masses. Computing the effects of the pairing properties of BDs on the
observed mass function, \cite{TK08} deduce that the IMF must be
discontinuous near $0.08\,M_\odot$.

Massive stars and BDs thus cannot be included in the BBP formalism
above, but require their own mathematical rules. In terms of physics
this means that the formation of massive stars and brown dwarfs occur
in a physical regime different to that of the typical star.

This is naturally understandable by noting that massive stars can only
form in the densest regions of their parent embedded cluster.  BDs, on
the other hand, typically form in the outer regions of extended
circum-stellar discs of the average star, either by disc instability
(\cite{GW07}) or by tidally-induced gravitational instability caused
by passing stars in the parent cluster (\cite{Tetal10}), or are
ejected embryos (\cite{RC01}).

In terms of numbers, massive stars and brown dwarfs are much rarer
than the typical star: on average one BD forms per five late-type
stars, while one massive star forms per few hundred late-type stars. 

\section{IBP mass regimes and IMF structure}
\label{sec:IBP-IMF}

\noindent
A possibly interesting issue thus emerges: While the BD/star schism is
clearly evident in the different pairing rules and the discontinuity
in the IMF near $0.08\,M_\odot$, the change of pairing rules between
late-type and early-type stars (roughly at a few~$M_\odot$) does not
seem to be evident in any corresponding structure in the IMF. In fact,
the flattening of the IMF near $0.5\,M_\odot$ does not seem to
correspond to a change in birth binary-star properties. While this
fact does not contradict the above statements, it does provide an
additional constraint on star-formation theories.

\section{Dynamical Population Synthesis (DyPoS)}
\label{sec:dypos}

\noindent
Having thus obtained a formal mathematical description of the
invariant birth binary population it now becomes possible to perform
{\sc Dynamical Population Synthesis} (DyPoS).  DyPoS rests on the same
ansatz as has already been applied to model the vertical structure of
the Milky Way disk (\cite{K02}) and to model the stellar initial mass
function of a whole galaxy (\cite{KW03}) by adding up the
contributions by each embedded cluster. This {\sc Lego Principle}
comes about from the realisation that stars that ultimately end up
populating the field form in groups embedded in gas
(e.g. \cite{LL03}).

If the birth stellar population in the groups {\it and} the
transformations that occur within the groups before the population
hatches into the field are understood and are known, then an entire
galaxy can be synthesised.  This principle is easy to implement for
the integrated initial mass function (IGIMF). It is a little bit more
involved when implementing it for an integrated description of the
vertical structure of the Milky Way disk because the stellar velocity
field emerging from an embedded stellar group needs additional
assumptions on its time-evolving properties.

Along the same lines, the binary population in a galaxy follows from
adding up all the binary populations born in the embedded groups
taking into account the dynamical transformation that acts on the
population in each group as it evolves and disperses,
\begin{equation}
\begin{split}
{\cal D}_{\rm field}(lP, e, q : m_1) = & 
\int_{M_{\rm ecl,min}}^{M_{\rm ecl,max}} \int_{R_{0.5,min}}^{R_{0.5,max}}
\Omega^{M_{\rm ecl},R_{0.5}} \left[ \Omega_{\rm
    eigenevol}\left[ {\cal D}_{\rm birth}(lP, e, q : m_1)\right] \right] \\
& \times \xi_{\rm ecl}(M_{\rm ecl},R_{0.5}) \; dR_{0.5}\,dM_{\rm ecl},
\end{split}
\label{eq:krdypos}
\end{equation}
where $\xi_{\rm ecl}(M_{\rm ecl}, R_{0.5})$ is the distribution of
stellar masses and half-mass radii of embedded aggregates of stars
(``embedded clusters'') forming in a time interval $\delta t$
($\approx 10\,$Myr) throughout the galaxy. Eq.~\ref{eq:krdypos} sums
up all the evolved populations of binaries that are formed in
$\xi_{\rm ecl}\,dR_{0.5}\,dM_{\rm ecl}$ clusters. $R_{0.5}\approx
0.4\,$pc for embedded clusters and $\xi_{\rm ecl} \propto M_{\rm
  ecl}^{-\beta}$ is the power-law mass function ($\beta=2$ is the
usually found index from observational surveys).

Two examples of DyPoS have already been computed by Michael Marks at
Bonn University. Eq.~\ref{eq:krdypos} is solved on a grid of $R_{0.5}$
and power-law index, $\beta$, of the embedded group or cluster initial
mass function, $\xi_{\rm ecl}(M_{\rm ecl})$, where the maximal
star-cluster mass, $M_{\rm ecl,max}$ follows from the
star-formation-rate (SFR) versus $M_{\rm ecl,max}$ relation of
\cite{WKL04}.  In an elliptical galaxy which formed in a burst with
$SFR=10^4\,M_\odot/$yr the most-massive star-cluster weighs about
$10^8\,M_\odot$ corresponding to the mass-scale of ultra-compact dwarf
galaxies. Binary systems are disrupted efficiently in the massive
clusters. Using Eq.~\ref{eq:krdypos} and assuming the typical embedded
star cluster has a radius of 0.4~pc, the binary fraction of the
late-type stellar population is computed to be $f_{\rm bin}=0.45$.  In
a dIrr galaxy with a $SFR=10^{-4}\,M_\odot/$yr the most massive
cluster that can form has a mass of about $M_{\rm
  ecl,max}=100\,M_\odot$. For $R_{0.5}=0.4\,$pc DyPoS yields $f_{\rm
  bin}=0.85$, because many more wider binaries survive the on average
less-massive embedded clusters in the dIrr galaxy, compared to the
above star-burst E galaxy.

A full mathematical treatment is found in \cite{MOK10}.

\section{Conclusions}
\label{sec:concs}

\noindent
It appears that the star-formation outcome in terms of stellar masses
and multiple systems can be formulated by the {\sc Star Formation
  Universality Hypothesis} (\S~\ref{sec:sfuniv}). For stars with
$m\simless 2\,M_\odot$ the {\sc Birth Binary Population}
(\S~\ref{sec:IBP}) can be defined. This is the outcome of star
formation in low to intermediate density ($\rho\simless
10^4\,M_\odot$/pc$^3$) cloud regions (e.g. of an embedded
cluster). For $m\simgreat 2\,M_\odot$ stars the pairing rules change
(\S~\ref{sec:IBP2}) perhaps reflecting the outcome of star formation
in dense regions such as in the cores of embedded clusters
($\rho\simgreat 10^5\,M_\odot$pc$^3$). Brown dwarfs follow entirely
separate rules (\S~\ref{sec:IBP2}) being an accompanying but distinct
population to stars. It remains to be understood why these changing
IBP properties do not correspond to the structure evident in the IMF
(\S~\ref{sec:IBP-IMF}).

\vspace{3mm}
\begin{footnotesize}
\noindent{\bf Acknowledgments:} I thank the organisers for a
stimulating and memorable conference in Barcelona.  My warmest
gratitude I wish to express to Sverre Aarseth for his brilliant work
on Nbody codes which are freely available and without which this work
would not have been possible, and for his unwavering support. This
text was written while being a Visitor at ESO/Garching, and I am
thankful for the kind hospitality of my colleagues there.
\end{footnotesize}


\end{document}